\documentclass[prb,aps,preprint]{revtex4}
\usepackage{amssymb}
\usepackage{amsmath}
\usepackage{graphicx}
\DeclareGraphicsExtensions{.eps}

\begin{document}

\bibliographystyle{apsrev}
\title{Field-induced charge-density-wave transitions
in the organic metal $\alpha$-(BEDT-TTF)$_2$KHg(SCN)$_4$ under
hydrostatic pressure}
\author{D.~Andres$^1$}
\author{M. V.~Kartsovnik$^1$}
\author{W.~Biberacher$^1$}
\author{K.~Neumaier$^1$}
\author{I.~Sheikin$^2$}
\author{H. M\"uller$^3$}

\affiliation{$^1$Walther-Meissner-Institut, Bayerische Akademie der
Wissenschaften, D-85748 Garching, Germany}

\affiliation{$^2$High Magnetic Field Laboratory, CNRS, F-38042
Grenoble France}

\affiliation{$^3$European Synchrotron Radiation Facility, F-38043
Grenoble, France}

\begin{abstract}

Successive magnetic-field-induced charge-density-wave transitions
in the molecular conductor $\alpha$-(BEDT-TTF)$_2$KHg(SCN)$_4$ are
studied in the hydrostatic pressure regime, in which the zero
field charge-density wave (CDW)state is completely
suppressed. The orbital effect of the magnetic field is
demonstrated to restore the density wave, while the orbital
quantization induces different CDW states in different field
intervals. In particular, we have found that at certain field
orientations the hysteretic first order transitions between
field-induced CDW subphases become visible at much higher
temperatures. This observation is very well in line with the
existing theories of field-induced CDW transitions.

\end{abstract}

\maketitle

\section{Introduction}

Phase transitions in low-dimensional molecular conductors induced
by rather high magnetic fields have become an intensively studied
topic during the last two decades \cite{ish98}. Among the most
prominent examples are the field-induced transition to a
spin-density-wave state \cite{dan96,ish98} or, recently
discovered, the field-induced superconductivity
\cite{uji01,uji03}. The former effect in strongly anisotropic
quasi-one dimensional (Q1D) electron systems has its origin in an
effective reduction of the dimensionality due to the orbital
motion of charge carriers in magnetic field on open
sheets of the Fermi surface \cite{cha96}, therefore being called
orbital effect.

In the layered organic metal $\alpha$-(BEDT-TTF)$_2$KHg(SCN)$_4$ there
exists a charge-density-wave (CDW) state below
8~K at ambient pressure \cite{bis98,chr00,ken97,fou03}. A Q1D
electron band becomes gapped at the Fermi level, due to
the so-called nesting of the Fermi surface, while  the other,
quasi-two-dimensional (Q2D) band still determines a metallic character
of the system.

It has been found, that hydrostatic pressure deteriorates the
nesting conditions and even leads to a complete suppression of
the density-wave at $P_0~\approx~2.5$~kbar \cite{and01,andres05}.
The suppression is easily explained by an increase in the
dimensionality of the Q1D band with hydrostatic pressure or
correspondingly an increase of the ratio between the effective
nearest and next-nearest interchain hopping integrals $t_c/t_c'$
within the conducting {\bf a-c} plane. The complete suppression
of the CDW state at $P_0$ has also been demonstrated to be
directly reflected in a distinct impact on another,
superconducting state also existing in this compound
\cite{andres05}.

Remarkably, it was shown that the orbital effect of magnetic field
exists in this CDW system \cite{and01}.
By applying a magnetic field, oriented along the least conducting
direction, it is possible that the imperfectly nested CDW state,
at P=1.5-2.5~kbar, even becomes stabilized before the suppression
by the additional, Pauli paramagnetic effect sets in.

Further, it was found that effects of orbital quantization take
place in the present compound \cite{and03}, where the nesting
vector appears to shift or jump between quantized values on
changing the magnetic field. Qualitatively, this effect emerges,
if the nesting conditions of the Fermi surface become so bad that
free carriers would start reappearing on the 1D sheets of the
Fermi surface \cite{ish98}. In the present case it was found that
within the high field CDW$_x$ state, existing above the
paramagnetic limit \cite{bis98,chr00,and03,har04}, the Pauli
effect takes this unnesting role \cite{and03}. This in turn
suggests that strong worsening of the nesting conditions by
hydrostatic pressure also realizes orbital quantization effects
\cite{zan96,leb03}.

The situation is fairly similar to the well known SDW systems of
the Bechgaard salts \cite{ish98}. In those compounds, all
carriers on the open sheets of the Fermi surface can be
considered to be completely gapped up to the critical pressure,
while above they would become free, that eventually completely
suppresses the density wave. In a magnetic field, best oriented
along the least conducting direction, it is then possible to
again stabilize the density wave \cite{gor84,mon88,kwa86}.
However, there will now be quantized values of the nesting vector
most preferable which lead to different SDW subphases
\cite{yamaji84a,gor84,mon85,yamaji87,cha96,ish98} in different
field ranges. On going from one subphase to another one the
nesting vector shifts to another quantized level. At low enough
temperatures this shift even turns to an abrupt jump that causes
the phase transition to change from the second to the first order.
Similar effects under hydrostatic pressure, namely field-induced
CDW (FICDW) transitions, have already been proposed to occur in
this organic CDW compound \cite{leb03}. However, besides some
hints noted in our previous reports
\cite{and01,and01b,kartsovnik04,kar04b}, these phenomena are still
lacking experimental proof.

In this work we give direct experimental evidence that first order
FICDW transitions indeed exist under pressure. This is
especially demonstrated by distinct hysteretic structures in the
magnetoresistance, at sweeping the magnetic field up and down. In
particular, it is shown that, by tilting the magnetic field towards
the conducting plane, it is even possible to shift the onset
temperature of the FICDW first order transitions to much higher
values. This observation is shown to be in line with recent theoretical
models of the FICDW phenomenon.

\section{Experiment}

We have performed standard four point measurements of the
interlayer resistance. The typical sample resistances at room
temperature were $\sim 10^3-10^4~\Omega$ with contact resistances
of $\sim30 \Omega$. Overheating of the samples was always checked
to be negligible by choosing appropriate measuring currents, that
e.g. turned out to be as low as 100~nA at 100~mK.

To apply pressure, a big (\O=20mm) and a small (\O=10mm) BeCu clamp
cells were used. The pressure at low temperatures was
determined from the resistance of a calibrated manganin coil to
an accuracy better than $\pm$100~bar. The temperature was
monitored by the resistance of a RuO sensor below 0.3~K. The big
cell was mounted on the cold finger of a home made dilution
refrigerator, the sample being oriented so that its conducting
ac-plane was perpendicular to the magnetic field generated by a
superconducting magnet. In order to keep the lowest operating
temperature of 100~mK, the rate of the field sweeps were chosen
as small a 2~mT/sec. Further, at the lowest temperatures weak
demagnetization effects of the pressure cell became significant
and had to be taken into account at controlling the temperature.
All in all, the lowest temperature could be kept constant during
a field sweep up to 15~T to an accuracy of $\leq 10\%$.

Effects of field orientation were studied in the 28~T resistive
magnet at the HMFL in Grenoble using the small pressure cell. The
cell was mounted on a $^3$He two-axes rotation insert. The absolute
values of both angles determining the sample orientation could be
determined to an accuracy better than 0.5$^\circ$, and changed with
the resolution better than 0.05$^\circ$. Field sweeps at fixed field
orientations were made at temperatures down to 0.4~K. The
angle-dependent magnetoresistance at fixed field intensities was
measured by sweeping the polar angle $\theta$ at different azimuthal
angles $\varphi$. At reasonable sweep rates of $\sim
0.1^\circ/$sec the lowest achievable temperature was 0.7~K.

\section{Results and Discussion}

\subsection{The re-entrant CDW state under pressure}


The critical pressure $P_0$, at which the zero-field
density-wave transition becomes fully suppressed has been
determined as 2.5~$\pm$~0.1~kbar \cite{and01,andres05}. Above
$P_0$ we expect the CDW state only to become stabilized via the
orbital effect of magnetic field. Fig.~1 shows magnetic field
sweeps up to 15~T, with the field directed perpendicular to the
conducting plane, at 100~mK for different pressures covering the
whole pressure range investigated within this work. The data
presented in Fig.~1 are obtained on one and the same sample and
have been qualitatively reproduced on another one measured at the
same time. Since the pressures were applied successively, i.e.,
without opening the clamp cell, the magnetic field orientation is
exactly the same for each pressure.

One of the basic characteristics of the ambient-pressure CDW state
of the present compound is the strong magnetoresistance,
$R(10~T)/R(0~T) \sim 10^2$ at low $T$, most likely caused by a
reconstruction of the closed orbits of the Q2D carriers in
the presence of the CDW potential \cite{kar93,ken96}. At $\approx
11$~T the magnetoresistance has a maximum followed by a negative
slope associated with a reentrance to the closed orbit topology
due to magnetic breakdown
\cite{ken96} between the strongly warped open sheets of the Fermi
surface. The latter also allows the fast Shubnikov-de Haas (SdH)
oscillations at frequency $F_\alpha~=~670$~T corresponding to
the undisturbed Q2D band
to appear. Moreover, it is known, that there is an anomalously
strong second harmonic signal at $2F_\alpha$ as well as
additional SdH frequencies at $F_\lambda=170$~T and
$F_\nu=F_\alpha+F_\lambda$, which only appear in the CDW state.
The origin of these unexpected
 frequencies is still under debate \cite{kar93,hou96b,har99b,har01osc}.

Under pressure, the magnitude of the magnetoresistance in Fig.~1
becomes smaller that is related to the gradual suppression of the
CDW energy gap. Besides this, the curves show other
pressure-induced changes, in particular on crossing the critical
pressure $P_0$. Most significantly, at pressures $P \gtrsim P_0$
slow oscillations emerge in the magnetoresistance background.
With increasing pressure these oscillations gradually move up in
field that is visualized by the dashed lines in Fig.~1. Their
amplitude is maximum at 3-3.5~kbar and reduces at
further increasing pressure. Obviously, these slow oscillations
occur exactly in the pressure range, in which FICDW transitions
are expected, i.e. at $P>P_0$.

Another distinct change on crossing the critical pressure is the
field at which the fast oscillations start to become visible.
While at $P<P_0$ these oscillations appear at rather high fields,
shortly before the  magnetoresistance background reaches the
maximum, i.e. 5-7~T, they already start below $2$~T at $P>P_0$.
This is directly seen in Fig.~2, where  the field sweeps around
2~T are shown in an enlarged field scale at pressures above and
below 2.5~kbar.

To better understand these changes at $P>P_0$, it turns out to be
useful to take a closer view on how the slow oscillations develop
on lowering the temperature. In Fig.~3 field sweeps taken at
3~kbar are shown for different temperatures. At 4.2~K the
resistance increases rather moderately with field and no
sign of any anomaly is seen. We therefore consider the normal
metallic state at this temperature to be present over the whole
field range. At 2.5~K, a stronger enhancement of the
magnetoresistance starting from $\approx 6$~T indicates the
reentrance into the CDW state. The orbital effect establishes the
density-wave state. With lowering the temperature the enhancement
of the magnetoresistance shifts to lower fields. Remarkably, the
slow oscillations only appear in the field region where the
magnetoresistance is elevated. This strongly suggests
that the slow oscillations only exist within the re-entrant CDW
state.

Within the whole temperature range the slope of the
magnetoresistance below 2~T remains approximately the same.
Moreover, in this field and pressure range the resistance turns
out to be even nearly temperature independent below 1~K as can be
seen from Fig.~4, where the field sweeps at 0.1 and 1~K are
shown at $P=3.5$~kbar. This coincides very well with the previous
observation that the normal metallic state exists at low fields at
$P>P_0=2.5$~kbar \cite{andres05}. The cylindrical orbits on the
Fermi surface of the Q2D carriers are then undisturbed, i.e. no
breakdown gaps exist anymore. Hence, it becomes clear why the
$\alpha$-oscillations start at such low fields, as shown in
Fig.~2.

The presence of the CDW state at higher fields at $P>P_0$ is
directly reflected in its distinct properties: First, in the
field range of 10-15~T the additional SdH frequencies $F_\lambda$
and $F_\nu$, characteristic of the CDW state, are observed. An
example of the Fast-Fourier-Transformation (FFT) spectrum of the
magnetoresistance at 3.5~kbar is given in Fig.~5. Surprisingly,
the frequency $F_\lambda$ is found to be pressure independent,
unlike $F_\alpha$ which in our studies shows a pressure
dependence of 17~T/kbar.
Second, there is a broad hysteresis in the magnetoresistance
between up- and downward sweeps of the magnetic field at $B \geq
3$~T. In Fig.~3 up and down sweeps of the magnetic field are
plotted for the lowest temperature, where the broad hysteresis is
clearly seen. Such a hysteresis is definitely inconsistent with  a
normal metallic behavior. On the other hand, it is known to be
present in the CDW state of this compound \cite{sasaki93a}.
Third, on lowering the temperature a strong decrease of the
magnetoresistance background is observed at $B~\gtrsim~8$~T, as
can be seen in Fig.~4. This is combined with a phase inversion of
the fast $\alpha$ oscillations as marked in Figs.~3 and 4 for 3
and 3.5~kbar, respectively, by vertical dashed lines. Such a
behavior has already been found to occur deep in the CDW
state and was discussed in a number of publications
\cite{kar97,honold99,har00,and01b}.

Altogether, the reentrance to the CDW state in magnetic field at
pressures between 2.5 and 4~kbar is clearly seen in the
magnetoresistance data. The measured phase transition fields and
temperatures are qualitatively well described by the theoretical
{\it B-T} phase diagrams of a Q1D CDW system at different nesting
conditions, which were proposed by Zanchi et al. \cite{zan96}.

Now we turn to the origin of the slow oscillations which exist
only in the re-entrant CDW state. At first glance one can suppose
that these are SdH oscillations emerging due to small pockets on
the Fermi surface, induced by imperfect nesting. This would give a
SdH signal of a very low frequency in $1/B$. Indeed, a FFT
transformation in the whole, inverted, field range within the
re-entrant CDW state shows a peak at about 20~T. The spectra of
the oscillations given in Fig.~1 are shown in Fig.~6. Since these
peaks are deduced from only very few oscillation periods it is
hard to judge about their exact positions. Moreover, since the
magnetoresistance background in the CDW state is not known and was
evaluated by a low order polynomial fit, an artificial shift of
the peak positions $\leq 1$~T might arise in the FFT spectrum. We
therefore cannot judge about the pressure dependence of the low
frequency. Nevertheless, a periodicity of these oscillations in
$1/B$ is clearly reflected. However, as will be pointed out
below, there are several observations which contradict the theory
of the normal quantum SdH oscillations, and favor the existence
of FICDW transitions.


\subsection{Field-Induced CDW Transitions at Perpendicular Field}


Fig.~7 shows the magnetoresistance obtained from the up-
($R_B^\uparrow$, black curve) and downward
($R_B^\downarrow$, grey curve) sweeps of magnetic field at
$P=3$~kbar, $T=0.1$~K. The raw magnetoresistance at increasing
field is shown in the figure by the dotted grey line. The lower
curve in Fig.~7 shows the difference
$\Delta$R=R$_B^\downarrow$-R$_B^\uparrow$ between the up and down
sweeps, demonstrating a considerable hysteresis, that was already
mentioned above in Sec.~IIIA. The hysteresis exhibits a clear
structure correlated with the slow oscillations of the
magnetoresistance background: its maxima are located at
approximately the field values corresponding to the maximum
curvature in R$_b(B)$.

Another peculiarity of the slow oscillation is the temperature
dependence of the phase. To illustrate the temperature dependence
of the phase of the slow oscillation, dashed lines following the
maximum curvature of the slow oscillations are placed into Fig.~3.
This anomalous behavior is certainly not expected for normal SdH
oscillations. On the other hand, the curves themselves, however,
are qualitatively quite similar to those observed in the FISDW
states of the Bechgaard salts \cite{kajimura83,kornilov02}.

 There are further similarities to the FISDW transitions, such as
for example the pressure dependence of the transition fields shown
in Fig.~8. The FICDW transition fields were defined from Fig.~1 as
the fields of maximum curvature of the magnetoresistance
background. Such a choice looks reasonable since these points
also correspond to the maxima in the hysteresis structure at
3~kbar. The obtained transition fields at 100~mK move
approximately linearly to higher values with enhancing the
pressure.
Note, however, that this shift with pressure is quite strong. For
SdH oscillations, this would correspond to an expansion of the
Fermi surface orbit area by about 20$\%/$kbar. The resulting
 increase of the SdH frequency must therefore be clearly
resolved in the FFT spectrum. Since this is not the case here,
this gives another argument against the usual SdH effect as the
reason for the observed slow oscillations. On the other hand, a
shift of the FISDW transitions to higher fields with increasing
pressure, or increasing $t_c'$, has already been demonstrated
experimentally \cite{kang93} on the Bechgaard salts. Thus, the
observed behavior is consistent with what one would expect for
FICDW.

In the whole, we regard these observations as a sign of first
order transitions between the FICDW subphases.
Unlike the sharp, well defined hysteretic FISDW transitions
\cite{kornilov02}, the transitions in our compound as well as the
peaks in the hysteresis are found to be smeared.
 The hysteretic first order transitions obviously appear only
at such low temperatures, $T\sim 0.1$~K. Hence, relative to the
ambient pressure density-wave transition temperature, these first
order FICDW transitions exist in a much lower temperature range in
comparison to the
 known FISDW cases of the Bechgaard salts.
Indeed, this is exactly what has been predicted by Lebed
\cite{leb03}. The reason for this comes from the fact that in the
CDW the nesting vector couples carriers of the same spin band
\cite{solynom79,gru94,ken97}. That effectively causes the
paramagnetic suppression of the CDW at high fields
\cite{fulde64,buz83b,zan96,grigoriev05} and also has to be taken
into account in the FICDW regime. Actually, the quantization
condition of the nesting vector must be extended by an additional
Zeeman or Pauli term \cite{leb03}:
\begin{equation}
\begin{split}
Q_{x}= 2k_F \pm q_{x,\text{Pauli}} + q_{x,\text{orb}} = &2k_F \pm
\frac{2 \mu_B B}{\hbar v_F} + N \cdot
G;~~~N=0,\pm1 \pm2 ... \\
&G=\frac{2ea_yB_z}{\hbar}
\end{split}
\label{fullquantequ}
\end{equation}
$Q_{x}$ being the nesting vector component in the conducting chain
direction, $k_F$ the Fermi wave vector, $\mu_B$ the Bohr magneton,
$v_F$ the Fermi velocity of the Q1D part of the electron system,
a$_y$ the lattice parameter perpendicular to the conducting chains
within the layer and $B_z$ the field component perpendicular to
the conducting planes. The right hand side of Eq.~(1) is
equivalent to two sets of quantized levels, one for each spin
subband. If the quantized values for both spin bands do not match
each other, the effective CDW coupling constant decreases, and so
does the transition temperature of the FICDW state, as well as
the onset temperature of the first order transitions.

Finally, we note that a modulation of the SdH oscillation amplitude
of the $\alpha$ frequency in the FICDW states is observed at the
lowest temperature (see Fig.~1). Its nature, however, is unclear at
present. No direct correlation between the modulation and the FICDW transitions
has been found so far. Further measurements are needed to draw any
reliable conclusions.

The discussion of Eq.~(1) showed that the mismatch of the
quantized levels for different spin bands is the reason for the
suppression of the onset temperature of the first order transitions.
As we will show next there is a possibility to enhance the density
wave instability, by changing the magnetic field orientation.

\subsection{FICDW transitions at different orientations of magnetic field}

An important consequence of tilting the field by an angle $\theta$
from the perpendicular direction towards the conducting plane is the
following:
 while the Zeeman spin splitting can be considered as an isotropic
effect, the orbital quantization depends only on the field component
perpendicular to the conducting planes, $B_z=B\cos\theta$. On tilting
the field the quantized levels of each spin subband therefore move
closer to each other but the distance between the $N=0$ levels of the
subbands, $\mu_BB/\hbar v_F$, remains the same. At certain angles
one, therefore, expects the quantized spin-up levels to coincide with
the spin-down ones. For such ''commensurate splitting'' (CS) angles
 the re-entrant CDW is predicted to become stabilized
at higher temperatures \cite{bjelis99,leb03}.

In Fig.~11 the scaled field sweeps for
different angles, covering a wide angular range, are shown.
Obviously, the amplitude of the slow oscillations is
strongly angle dependent.
In this angle interval, 0$^\circ$ - 74$^\circ$, there are
two regions, around 57$^\circ$ and 71$^\circ$, where the amplitude
of the slow oscillations has a maximum, whereas around 43$^\circ$
and 65$^\circ$, it nearly vanishes.
Besides this angular oscillation of the amplitude, the transition
fields also provide some information: on crossing the angle where
the amplitude nearly vanishes, the transition fields shift by half a
period. The dashed lines in Fig.~11  mark the transition fields,
where the oscillation has a maximum curvature. As can be seen, the
transition fields change several times their positions on tilting the
magnetic field. Thus, the slow oscillations are found to possess
some kind of "spin zeros" at certain field directions, as known from
the normal SdH effect. In the latter effect the phase of the
oscillations inverts several times on tilting the magnetic field
\cite{sho84,wos96}.

To understand this behavior in the present case of field-induced CDW
transitions, we first have to recall in a qualitative manner what
happens in a SDW system. In the FISDW pressure range there are
preferable, quantized values of the $x$ component of the nesting
vector. The spin susceptibility or response function $\chi(Q_x)$ of
the system therefore is a more or less periodic function with maxima
at such values of the nesting vector \cite{mon85}. At not too low
temperatures we assume this response, expanded into a harmonic
series, to be strongly dominated by its first harmonic. If we now
switch to the CDW system, one has to consider the Zeeman energy
splitting of the different spin bands.


Assuming the response in the CDW system of each spin band also to be
well described by its first harmonic, the situation resembles the
 Landau quantization. The superposition of both response
functions gives a total oscillating response function, where the
phase keeps the same at a constant out-of-plane field component,
$B\cos \theta$, independent of the field orientation.
However, the sign of the oscillation may change, that determines a
phase inversion. The same mechanism leads to the spin zeros in the
quantum SdH and de Haas-van Alphen effects. This means that since
the orbital quantization only depends on the magnetic field
component in the least conducting direction, there is the
possibility that by tilting the magnetic field the oscillation of the
response function inverts its phase. Since the quantized values are
determined by the maxima of the total response function, these
quantized values will shift by half a period. Hence, one can also
expect the FICDW transitions to shift in magnetic field. Noteworthy,
we do not expect the same effect to happen in SDW systems, since in
the latter the nesting vector is not influenced by the Pauli effect.

In the picture described above, a change of the FICDW transition fields
 should happen at exactly those angles at which quantized
values of the nesting vector, Eq.~(1), for one of the spin subbands
lie exactly in the middle between those of the other spin subband.
These angles are easily derived to be determined by:

\begin{equation}
\cos(\theta)=
\frac{1}{\mathrm{M}+0.5}\left[\frac{2\mu_B}{v_Fea_c}\right];
 \text{ M=$\pm1,\pm2...$}. \label{spinzero}
\end{equation}

Another interesting observation is shown in Figure~9, where
magnetic field sweeps, with the field scaled in
$\cos\theta$, are shown at different angles in the narrow
interval $\theta=50^\circ-60^\circ$ at 2.8~kbar. The black curves a
re taken
on sweeping the field up, grey curves on sweeping down. Obviously,
the slow oscillations do scale in $\cos\theta$. This is expected,
since the quantization is determined by the field component
$B_z=B\cos(\theta)$.
As in the case of the perpendicular oriented field at 3~kbar,
see Fig.~7, a modulated hysteresis between both sweep directions with
rather sharp peaks is observed. This can be directly seen in Fig.~10
where the magnitude of the hysteresis is plotted against the field
for different tilt angles. Within this angular range this hysteresis
becomes strongest at the field-induced transition marked by the
dashed line in Fig.~9. Further, it is seen from Fig.~10 that at
57.7$^\circ$ the magnitude of this hysteresis has a maximum. Note
that the temperature here, $T=0.45$~K, is much higher than for the
data in Fig.~7; at perpendicular field direction no structure in the
hysteresis, corresponding to the different FICDW transitions, has
been resolved.

The fact that this hysteretic structure was only observed in this
narrow angular range, makes us believe that this is an effect of
the first CS angle. As mentioned at the beginning
of this section, at the CS angles the total CDW susceptibility
becomes stronger and the coupling constant as well
as the transition temperature of the density wave should increase
\cite{leb03}.

Thus, altogether, if we know the angles where the phase of the slow
oscillations changes, there is then the possibility to directly
evaluate the CS field orientations, i.e. the directions, where
the first order transitions should appear at higher temperatures.
These must be given by \cite{bjelis99,leb03}:

\begin{equation}
\cos(\theta)=
\frac{1}{\mathrm{M}}\left[\frac{2\mu_B}{v_Fea_y}\right].
\label{commensdir}
\end{equation}

In our experiment many field sweeps were performed at different
angles. Since the angular step was chosen rather small, the first
three ranges, in which the sign reversal of the slow oscillation
occurs, are found to be restricted to the narrow intervals:

$\theta_{S1}=40^\circ-45^\circ$, $\theta_{S2}=63.7^\circ-65.7^\circ$ and
$\theta_{S3}=73^\circ-73.9^\circ$.

These angular intervals are in good agreement with the assumed
1/$\cos(\theta$) scaling, Eq.~(2), as can be seen in Fig.~12 where
the commensurability index is plotted against 1/$\cos(\theta)$. The
corresponding Fermi velocity amounts to
$\approx$~$1.2\cdot10^5$~m/sec, that is about twice the value
determined at ambient pressure by Kovalev et al. \cite{kov02}:
$0.65\cdot 10^7$~cm/sec.


From Eq.~(3) and Fig.~12 the first
CS angle may thus be evaluated to:
$\theta_{c1}=55.60^\circ-58.97^\circ$; i.e. exactly the angular
range where we find a pronounced increase in the hysteresis,
see Fig.~10. This remarkable agreement between our experimental
results and this simple model above gives another strong argument
for the existence of FICDW subphases at $P \geqslant P_0$.

Noteworthy, from Fig.~12 it follows that at perpendicular field
direction we are by far not at a CS angle, so that here the
coupling constant must be rather low. This explains why at this
field orientation the hysteretic first order transitions appear at
much lower temperatures than at the CS angles.

\subsection{Angle dependent magnetoresistance}

Now that the general behavior under pressure is qualitatively
understood, we discuss the effects which can be seen in the
angle-dependent magnetoresistance oscillations (AMRO) at $P>P_c$.

The field orientation is always
turned in a plane normal to the layers and the angle $\theta$ is
determined as before.
The measurement of AMRO is a powerful tool to determine the Fermi
surface in such low-dimensional electron systems
\cite{kartsovnik04c}.
For the present compound the AMROs in the CDW state
well describe strongly warped open orbits, which are
effectively created by the Q2D electrons being exposed to the CDW
potential \cite{kar93}. They show
pronounced dips, periodic in $\tan\theta$, in the otherwise
rather high magnetoresistance. We will refer to this kind
of AMRO as the ''Q1D type''.
By contrast, in the NM state the
AMRO are of the ''Q2D type'', i.e. mostly determined by the cylindrical part of
the Fermi surface, showing pronounced peaks in the
magnetoresistance, also periodic in $\tan(\theta)$. Therefore, a
field-induced transition from the NM to the CDW state should also be
resolved in AMRO measurements.

A set of AMROs measured at 2.8~kbar and 0.7~K is shown in Fig.~13.
The superposed fast oscillations developing at $B \geq 6.5$~T are
SdH oscillations. These curves were measured in the same cooling run
as the data shown in Figs.~9 and 11, i.e., the pressure
was exactly the same. At 1~T, the AMRO, although very
weak, shows a normal metallic behavior, as expected from our
discussion above. On enhancing the field, see 4.3, 6.5 and 9~T curves in
Fig.~13, the magnetoresistance is found to increase strongly
at low angles. In effect, at the border of this low-angle
region rather sharp dips develop, as is marked in Fig.~13 in the 9~T
sweep by arrows. A clear Q2D-type of the AMRO is now only observed
at higher angles, $\theta \gtrsim 70^\circ$, while in the
mid-angular range the modulation of the magnetoresistance appears to
be more complex. On increasing the field further, 15~T and 20~T, the
low-angle hump gradually fades and the curves again become Q2D like.

We start with the pronounced hump in the low-angular range. Such a
behavior is definitely not expected for the NM state but is fully in
line with our proposal that here the CDW is established due to the
orbital effect. This means that the strong magnetoresistance at low
tilt angles as well as the dips, marked by the arrows in Fig.~13,
originate from the Q1D-type AMRO. The change to the Q2D-type AMRO
above $70^\circ$ we interpret by the reduction of the orbital
effect,since the latter only depends on the perpendicular field
component. On
tilting the field the CDW energy gap becomes smaller, and with it
the magnetic breakdown gap. Therefore, the Q2D-type AMRO must become
dominant.

At higher fields, 15~T and 20~T, strong magnetic breakdown sets in,
so that even at low angles the AMRO again becomes dominated by the
Q2D Fermi surface. This is directly reflected in the quite strong SdH
oscillations which are superposed on these curves. Such a behavior is already well
known from ambient pressure measurements, where a clear crossover
 from the Q1D AMRO regime to the Q2D AMRO regime occurs with field
due to the magnetic breakdown \cite{hou96c}.

Altogether, the measured AMROs are thus consistent with the
results discussed above. Since the hysteretic first order transitions were
found to appear already at higher temperatures at the CS field
orientations one could also expect some additional
features to emerge at these angles. As we have seen in
Fig.~9 the first CS angle must be around 57.7$^\circ$.
Although there seem to be some anomalous features in between
$40^\circ-60^\circ$ in the plots shown in Fig.~13, there is
obviously no direct sign of this CS angle.
 This, however, is not too surprising, since the temperature
here is rather high, so that an effect of the CS angle already
smears out.

\section{Conclusion}

The presented results provide an evidence that at pressures
$P~\gtrsim~2.5$~kbar the CDW in $\alpha$-(BEDT-TTF)$_2$KHg(SCN)$_4$
only exists under magnetic field due to the orbital
effect. Within this re-entrant CDW state a slow oscillation in the
magnetoresistance background of very low frequency is observed. We
show that this oscillation displays the new phenomenon of FICDW
transitions. Direct evidence for this is given by a clear
hysteretic structure observed in the magnetoresistance at
perpendicular field direction at $P=3$~kbar and $T=100$~mK.
Moreover, the superposition of the orbital quantization and the
Pauli effect for different spin subbands is shown to cause a
modulation of the density-wave stability on tilting the magnetic
field orientation towards the conducting plane. This is
experimentally seen in a sign reversal of the slow oscillation
amplitude and, in particular, clear hysteretic structures at
elevated temperatures in the vicinity of the CS angle, where
the quantized values of the nesting vector for different spin
subbands superpose on each other. These observations demonstrate
the existence of FICDW transitions at $P>P_0$ in this organic
conductor.


\newpage

Fig.~1. Magnetoresistance at various
       pressures at 100~mK. Above $P_0~\approx$~2.5~kbar slow oscillations
       emerge in the magnetoresistance background. With increasing pressure these
       oscillations gradually move to higher fields as visualized by dashed
       lines.

Fig.~2. Expanded low field part of the curves from Fig.~1. Above 2.5~kbar the
       fast SdH oscillations start to appear already below 2~T.

Fig.~3. Magnetoresistance at $P=3$~kbar. The data are recorded at
        increasing field and different temperatures. The curves are
       offset from each other. At the lowest temperature the down
       sweep is additionally shown by the black curve. The dashed
       line marks a field, at which a maximum of the fast SdH
       oscillations turns at to a  minimum on lowering the
       temperature, that visualizes the phase inversion.

Fig.~4. Magnetoresistance recorded at two different temperatures and
        P~=~3.5~kbar. Note the temperature
       independent resistance at low fields, suggesting the NM state to
       be present. As in Fig.~3, the dashed line at higher fields
       illustrates the phase inversion.

Fig.~5. FFT spectrum of the magnetoresistance at $T=100$~mK and
$P$~=~3.5~kbar in the field interval 10-15~T. The additional
frequencies $\lambda$ and $\nu$ within the re-entrant CDW state
are clearly resolved.

Fig.~6. FFT spectrum of the whole measured field range; 2-15~T.
       The slow oscillations are reflected by an additional
       peak at $\approx 20$~T

Fig.~7. Up and down field sweeps of the magnetoresistance at
$P=3$~kbar and T$=100$~mK. The hysteresis, determined by subtracting
one curve from the other, is shown below. A clear structure matching
the oscillatory features can be seen.


Fig.~8. The FICDW transition points at 100 mK
       move to higher magnetic fields with increasing pressure.

Fig.~9. Field sweeps at different tilt angles of magnetic
        field, with the field scaled in $\cos(\theta$), at
        $T=0.4$~K, $P=2.8$~kbar. The curves at different angles are
        offset for clarity. Black curves show up sweeps of the
        magnetic field, grey curves down sweeps. The dashed line
        marks the position of the FICDW transition for which the
        observed hysteresis has a maximum.

Fig.~10. Hysteresis obtained from a subtraction of the up from
        the down sweeps of Fig.~9 at different field
        directions. The arrow points to the maximum in the hysteresis at
        the FICDW transition that is marked in Fig.~9 by
        the dashed line.

Fig.~11. Field sweeps at $T=0.4$~K, as in Fig.~9, in a bigger
        angular range. To illustrate the presence of ''spin zeros'',
        dashed lines are approximately placed at the extrema of the slow
        oscillations. On increasing the angle the oscillations
        show several times a change of the amplitude sign.

Fig.~12. Observed angular ranges of the spin zeros scaled
        according to Eq.~(2). The expected linear
        behaviour of the half odd integer quantum number in $1/\cos$
        allows an evaluation of the CS field directions at integer
        quantum numbers, Eq.~(3).

Fig.~13. Angle-dependent magnetoresistance oscillations for several
different fields at 0.7~K and 2.8~kbar. The curves are offset from
each other. At 1~T the resistance has been multiplied by 20. The arrows point to
the sharp dips, typical of the CDW state, in the curve at 9~T.


\end{document}